\documentclass[12pt,a4paper,final]{iopart}

\usepackage{iopams}
\usepackage{graphicx}
\usepackage[breaklinks=true,colorlinks=true,linkcolor=blue,urlcolor=blue,citecolor=blue]{hyperref}
\usepackage{color}
\begin{document}

\title{A polarization quantum key distribution scheme based on phase matching}

\author{Gao Feifei$^{1}$}
\address{$^1$Address One, College of Mathematics and Information Science, Shaanxi Normal University, Xi¡¯an 710119, China}

\ead{792099462@qq.com}

\author{Li Zhihui}
\address{Address Two, College of Mathematics and Information Science, Shaanxi Normal University, Xi¡¯an 710119, China}
\ead{lizhihui@snnu.edu.cn}

\author{Liu Chengji}
\address{Address Three,  Xidian University, Xi¡¯an 710119, China}
\ead{120148147@qq.com}
\author{Han Duo}
\address{Address Four, College of Mathematics and Information Science, Shaanxi Normal University, Xi¡¯an 710119, China}

\ead{1157389901@qq.com}

\begin{abstract}
The Quantum Key Distribution protocol can encode a single quantum state and implements an information-theoretically secure key distribution protocol in communication. In the actual QKD experimental system, there are usually two encoding methods which are phase encoding and polarization encoding. Ma et al.[Phase-Matching Quantum Key Distribution, Phys. Rev. X., 2018, 8£¨3)] proposed the phase-matching QKD protocol, which has high transmission and it is an extension of the measurement device independent QKD. This paper successfully gives the polarization scheme of this PM-QKD protocol, the bases in the polarization scheme are arbitrary, and eliminates detector side channel attacks. Furthermore, we give the security analysis and simulation results of the polarization scheme, and compare it with the BB84 protocol. The simulation results show that our protocol is superior to the BB84 protocol in terms of transmission distance under the fixed key rate.

\end{abstract}

\vspace{2pc}


\section{Introduction}

Quantum key distribution(QKD) technology is the main way to realize quantum secure communication at present, and it is very important to carry out research on it. In 1984, Bennett and Brassard proposed the first QKD protocol [1](BB84 protocol) . It encodes four polarization states of photons, whose polarization states can be divided into two groups of conjugated groups. Also, the two polarization states of each group are orthogonal to each other. For example, the two pairs of bases can be a horizontal vertical basis ${\{0^\circ,90^\circ\}}$ and a diagonal basis ${\{45^\circ,135^\circ\}}$, respectively. The specific process of QKD is as follows:
\par(1)Alice encodes the key information to be transmitted into the polarization state of the photon according to the encoding rule, and sends it to the receiver.
\par(2)Bob randomly selects the Z-basis or X-basis measurement for the received photons. Regardless of the measurement basis he uses, he is counted as 0 as long as he gets the first result of the Z or X basis; the second result is recorded as 1.
\par(3)When all photons are sent, Alice and Bob tell each other's respective bases sequences through the classic channel.
\par(4)They discard the data of different basis, then correction and secret amplification of the remaining data.
\par In the process of QKD protocol, some protocols and mersurement basis are needed to transport in the classical channel in order to assist the communication parties to obtain the final secure quantum key sequence. Therefore, the BB84 protocol has a risk that  the measurement basis is stolen in the classical channel. In 1992, Bennett et al. proposed the B92 protocol, which only needs to use two polarization states [2], and there is no need to ensure that they are orthogonal to each other. That process is similar to the BB84 protocol, but there is no basis step in the subsequent processing, so the risk of theft for the measurement basis can be avoided. However, since the receiver can only correctly receive $25\%$ of the results, which results in extremely low transmission efficiency.
\par Most of the light sources used in the above two polarization protocols are weak coherent states, rather than ideal single photon sources, resulting in photon number separation attacks(PNS) [3]. On the other hand, for photon detectors, there are time-shift attacks [4], faked states attacks [5], blind attacks [6], and so on. In response to these attacks, different improvement protocols were proposed. For example, for the imperfection of the light source, the decoy protocol [7-9] was proposed. Similarly, for the detector attack, the measurement device independent QKD(MDI-QKD)[10] was proposed by Lo et al. in 2012. The MDI-QKD protocol not only eliminates attacks on the detector side channel, but also has the ability to double the transmission distance. However, its key rate is still ${O(\eta)}$, where is the total transmission between Alice and Bob (ie the probability that a photon is successfully transmitted through the channel and detected). In order to improve the key rate, scholars have continuously improved the MDI-QKD protocol. In 2018, M. Lucamarini, Z.L. Yuan et al. proposed a twin-field QKD (TF-QKD)[11] protocol based on the phase MDI-QKD protocol [12]. The coding of the protocol is based on two sets of orthogonal basis vectors and is an extension of the BB84 protocol. The random phase is added to the protocol, so it can not only resist photon number separation attacks and detector attacks, but also increase the key rate from ${O(\eta)}$ to ${O(\sqrt{\eta})}$. However, the security of the TF-QKD protocol has not been proven. Ma Xiongfeng et al. proposed a phase-matching QKD (PM-QKD) [13] protocol which is immune to all possible measurement attacks in 2018, and proof of security are given based on an optical mode.
\par  From the above analysis protocols, the BB84 protocol, the B92 protocol, and the  MDI-QKD protocol are all belong to polarization coding protocols, but the TF-QKD and PM-QKD protocols are a new phase-coded MDI-QKD protocol. It can be known from the Ref [14] that phase encoding has the following advantages good stability, strong anti-interference ability, low bit error rate, etc., so it is widely used in optical fiber transmission. But the polarization state shift often occurs to cause a decrease in the interference contrast, resulting in a problem of an increase in the bit error rate. Thus, polarization coding has significant advantages in space transmission, it has become the first choice for free-space QKD experiments. However, in the transmission process of the optical fiber, the birefringence caused by the non-uniformity of the optical fiber during the drawing process, as well as the curvature of the optical fiber, the ambient temperature and the stress, etc., the photon polarization state is highly prone to irregular changes, which in turn destroys the original polarization state coded information of the photon when the bit error rate of the communication system increase. Therefore, studying the polarization QKD protocol based on phase encoding and the phase \textcolor[rgb]{1,0,0}{QKD} protocol based on polarization encoding have important significance both in theory and experiment.
\par A QKD scheme of the polarization state corresponding to the PM-QKD protocol is given based on PM-QKD protocol, and its security is analyzed in this paper. It is organized as follows: Section 2 reviews the PM-QKD protocol, and the phase-encoding polarization scheme is given in Section 3 and analyzes its security. The Section 4 gives the key rate simulation formula of our scheme and a comparison between the transmission distances of the two protocols in the case of a fixed key rate.

\section{PM-QKD protocol review}

In the PM-QKD protocol, the communicating parties Alice and Bob independently generate coherent state pulses. For a d-phase PM-QKD protocol, Alice and Bob encode their key information into the phase of the coherent state and send it to an untrusted measurement location that may be controlled by Eve, as shown in Figure 1. Eve is expected to perform interferometric measurements, defining successful detection as having one and only one click in two detectors, denoted as an L click or an R click. The following is the specific process of the PM-QKD protocol.
\par (1)State preparation-Alice randomly generates a key bit ${k_{a}}$ and a random phase ${\phi_{a}\in[0,2\pi)}$,and then prepares a coherent state${{|\sqrt{\mu_{a}}e^{i(\phi_{a}+{\pi}{k_{a}})}\rangle}_{A}}$. Similarly, Bob generates ${k_{b}}$ and ${\phi_{b}\in[0,2\pi)}$, then prepares ${{|\sqrt{\mu_{b}}e^{i(\phi_{b}+{\pi}{k_{b}})}\rangle}_{B}}$.
\par(2)Measurements-Alice and Bob send their light pulses A and B to an untrusted Eve, who is expected to perform an interference measurement and record the detector (L or R) that clicks.
\par(3)Announcement-Eve announced her detection results. Then Alice and Bob announce the random phases ${\phi_{a}}$ and ${\phi_{b}}$, respectively.
\par(4)Sifting-Alice and Bob repeat the above steps multiple times. When Eve announces a successful detection (just one detector L or R click), Alice and Bob make ${k_{a}}$ and ${k_{b}}$ the raw key bits. If Eve declares an R click, Bob flips his key bit ${k_{b}}$. Then, as long as ${|\phi_{a}-\phi_{b}|=0}$ or ${\pi}$, Alice and Bob's raw key are unchanged; when ${|\phi_{a}-\phi_{b}|=\pi}$, Bob flips his key bit ${k_{b}}$.
\par(5)Parameter estimation-Alice and Bob analyse the gain ${Q_{\mu}}$ and qubit error rates ${E_{\mu}^{Z}}$ from all of the retained raw data and then estimate ${E_{\mu}^{X}}$ using Eq(1) in Ref[15].
\begin{center} $$E_\mu^X\leq1-\frac{e^{-\mu}\frac{\mu^k}{k!}\times\frac{1}{2}(Y_{01}+Y_{10})}{\sum_{k=0}^\infty {P^\mu(k)Y_k}}.\eqno{(1)}$$
\end{center}
\par(6)Key distillation-Alice and Bob perform error correction and secret amplification on the filtered key bits to generate a private key.
\par Notations.--Denote a coherent state in mode A to be $|\sqrt\mu_a e^{i\phi}\rangle_A$, where $\mu$ is the intensity and $\phi$ is the phase, $\mu_a=\mu_b=\frac{\mu}{2}$, $k_{a(b)}\in{\{0,1\}}$  represents the key bit of Alice (Bob), total gain $Q_\mu$,  phase error rate ${E_{\mu}^{X}}$, and bit error rate ${E_{\mu}^{Z}}$.
\begin{figure}[htb!]
 \centering
 \includegraphics{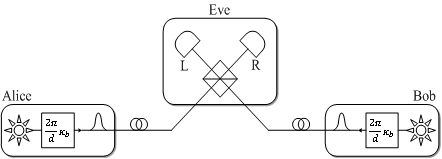}
 \caption{\label{figureone}Illustration of the PM-QKD protocol. }
\end{figure}
\par
The PM-QKD protocol passes the measurements to a third party, eliminating detector-side channel attacks. It is an extension of the MDI-QKD protocol, so the transmission distance is twice that of the BB84 protocol. The key rate of the protocol is the same as that of the TF-QKD protocol, which is $O(\sqrt\eta)$, which is the key rate that is not reached by other protocols[16-20]. Another advantage of the PM-QKD protocol is that there is no basis step, eliminating the classic communication between the two parties. Based on these advantages of the PM-QKD protocol, we give its polarization scheme.
\section{Polarization scheme based on phase encoding}
 This section gives our polarization scheme based on phase encoding. In the Section 2, we know that the coherent state of the PM-QKD protocol is ${{|\sqrt{\mu}e^{i(\phi_{a(b)}+{\pi}{k_{a(b)}})}\rangle}}$ and its phase is $\phi_{a(b)}+\pi k_{a(b)}$, where $\phi_{a(b)}\in[0,2\pi)$ is a random phase, and $\phi_a\neq\phi_b$.
\subsection{Specific process of the polarization scheme}
Let $\phi_a$ and $\phi_b$ are both random phases, and their corresponding bases are $M_1^\prime$, $M_2^\prime$ respectively, where $M_1^\prime$, $M_2^\prime$ are the bases represented by the polarization state. Thus, the polarization state corresponding to the random phase is the basis of the proposed scheme. The specific process is given below. Let $M_1^\prime=\{|\psi_{M_{11}}\rangle,|\psi_{M_{12}}\rangle\}$, $M_2^\prime=\{|\psi_{M_{21}}\rangle,|\psi_{M_{22}}\rangle\}$ be two different sets of bases in Bloch ball. The polarization scheme of the PM-QKD protocol is as follows.

\par(1)State preparation-Alice and Bob respectively prepare the polarization state of the photon and independently select the $M_1^\prime$ or $M_2^\prime$ basis encode the key information.
\par(2)Coding-If the key information transmitted by Alice(or Bob) is 0 ,she(or he) selects $M_1^\prime$ basis, the polarization state of the photon is $|\psi_{M_{11}}\rangle$; if she selects $M_2^\prime$ basis, the polarization state of the photon is $|\psi_{M_{21}}\rangle$;  if the key information transmitted by Alice(or Bob) is 1, she selects $M_1^\prime$ basis, the polarization state of the photon is $|\psi_{M_{12}}\rangle$; if she selects $M_2^\prime$ basis, the polarization state of the photon is $|\psi_{M_{22}}\rangle$.
\par(3)Measurement-Alice and Bob send the photon's polarization state to Eve, Eve receives the photon and measures it, recording the click detector (L or R).Eve measures the polarization states of Alice and Bob. If the phase difference corresponding to the polarization state is 0 then the click is L; and if the phase difference corresponding to the polarization state is $\pi$ then the click is R.
\par(4)Announcement-Eve announced the detector results, Alice and Bob announced their respective basis.
\par(5)Flip-Alice and Bob repeat the above steps multiple times. When Eve announces a successful detector click, Alice and Bob make $k_a$ and $k_b$ the raw key bits. If Eve declares an R click, Bob flips his key bits.
\par(6)Base step-Alice and Bob check their bases, then leaving the key bits of $M_1^\prime=M_2^\prime$.
\par As can be seen from the introduction of this paper that the bases used in the BB84 protocol and the MDI-QKD protocol are both Z-basis or X-basis, as shown in Figure 2, and the two sets of bases are special. In our scheme, two different sets of bases can be arbitrarily selected according to the value of the random phase. These two sets of bases are not necessarily orthogonal, which is the advantage of this scheme.
\begin{figure}[hbt!]
 \centering
 \includegraphics{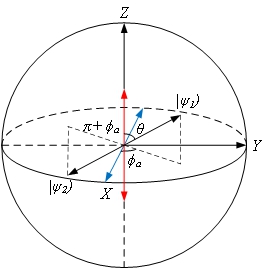}
 \caption{\label{figureone}The selection of the polarization scheme basis: the red vector in the figure is the Z basis, the blue vector is the X basis, and the black vector is the arbitrarily selected basis; it can be seen that the Z basis and the X basis are orthogonal to each other. }
\end{figure}

\par \textcolor[rgb]{1,0,0}{Here, Alice and Bob are sending coherent states of photons. The corresponding relationship is as follows: $|\psi_{M_{11}}\rangle\sim\sqrt \mu e^{i\phi_a}, |\psi_{M_{12}}\rangle\sim\sqrt \mu e^{i(\phi_a+\pi)}, |\psi_{M_{21}}\rangle\sim\sqrt \mu e^{i\phi_b}, |\psi_{M_{22}}\rangle\sim\sqrt \mu e^{i(\phi_b+\pi)}$}. \textcolor[rgb]{1,0,0}{Phase differences can be measured when the two communicating parties are sending coherent states.}

\subsection{Example}

First, we take $\theta=\frac{\pi}{2}$($\theta$ is the angle between the vector and the $Z$ axis), $\phi_a=\frac{\pi}{6}$ and $\phi_b=\frac{\pi}{4}$ in Figure 2. $|0\rangle$ and $|1\rangle$ qubit state are the two polarization of Z basis, which are horizontal polarization state and the vertical polarization state, respectively. Then the phase
$$ \pi k+\phi_a=\left\{
\begin{array}{rcl}
\frac{\pi}{6}, & &{k=0}\\
\frac{7\pi}{6},  & &{k=1}.
\end{array} \right.
$$
 Then the basis corresponding to $\phi_a$ is represented by the phase of the photon as the $M_1=\{\frac{\pi}{6},\frac{7\pi}{6}\}$. According to
\begin{center}
$|\psi\rangle=\cos\frac{\theta}{2}|0\rangle+e^{i\phi}\sin\frac{\theta}{2}|1\rangle$,
\end{center}
 we can write $M_1$ as polarization state $M_1^\prime=\{|\psi_{M_{11}}\rangle,|\psi_{M_{12}}\rangle\}$. Similarly, the basis phase corresponding to the random phase $\frac{\pi}{4}$ is represented by $M_2=\{\frac{\pi}{4},\frac{5\pi}{4}\}$, and the corresponding polarization state is $M_2^\prime=\{|\psi_{M_{21}}\rangle,|\psi_{M_{22}}\rangle\}$. Where
 \begin{center}
 $|\psi_{M_{11}}\rangle=\frac{\sqrt 2}{2}|0\rangle+\frac{\sqrt 2}{2}(\frac{\sqrt 3}{2}+\frac{1}{2}i)|1\rangle$,
 $|\psi_{M_{12}}\rangle=\frac{\sqrt 2}{2}|0\rangle-\frac{\sqrt 2}{2}(\frac{\sqrt 3}{2}+\frac{1}{2}i)|1\rangle$,
 $|\psi_{M_{21}}\rangle=\frac{\sqrt 2}{2}|0\rangle+\frac{\sqrt 2}{2}(\frac{\sqrt 2}{2}+\frac{\sqrt 2}{2}i)|1\rangle$,
 $|\psi_{M_{22}}\rangle=\frac{\sqrt 2}{2}|0\rangle-\frac{\sqrt 2}{2}(\frac{\sqrt 2}{2}+\frac{\sqrt 2}{2}i)|1\rangle$.
 \end{center}

\par In Section 3, the polarization scheme uses only two sets of bases. In fact, when $\phi_{a(b)}$ takes a value, it corresponds to a set of bases. The scheme of this paper is roughly the same as the whole process of the PM-QKD protocol, except that our scheme is belong to polarization coding and the other is belong to phase coding. Since the two values of the random phase difference of $\pi$ are in the same group basis, the phase sifting step of the original PM-QKD protocol is the basis step of the polarization scheme.

\subsection{Security analysis}

 The phase encoding protocol of PM-QKD is due to the publication of the random phase resulting in the failure of the tagging method used in the photon number channel model [21] and the Gottsman et al. security proof (GLLP security proof), so the proof of PM-QKD protocol is based on the optical mode [13]. After we change the phase encoding protocol of the PM-QKD protocol to the polarization scheme, and the random phase is the basis, then the tagging method in the photon number channel model and the GLLP security proof does not fail. It is essentially an extension of the MDI-QKD protocol. So, the security analysis for our scheme with decoy states follows from that of GLLP formula[22], which rely on the photon-number channel used in Ref[21]. The key rate for the our scheme is given by Eq(2).
 \section{Simulation}
 In this section, we use the parameters of Table 1 to simulate the performance of this paper. The comparison between the scheme and the BB84 protocol is given, and the simulation key rate formula are presented, respectively.

\subsection{Simulation formula for polarization scheme}
The key rate formula for the polarization scheme of this paper is based on the Ref[23]:
\begin{center}
$$R_{polarization}=\frac{1}{2}\{Q_{1,1;M_1^\prime}[1-H(e_{1,1;M_2^\prime})]-Q_{M_1^\prime}f(E_{M_1^\prime})H(E_{M_1^\prime})\} \eqno{(2)}$$,
\end{center}
\par Here,$Q_{n,m;M_1^\prime},Q_{n,m;M_2^\prime},e_{n,m;M_1^\prime}$ and $e_{n,m;M_2^\prime}$ represent the gain and qubit error rate of the signals transmitted by Alice and Bob, respectively. Where $n\ and\ m$ represent the number of photons sent by both communicating parties; $M_i^\prime (i=1,2)$ represent the choice of their basis. We selected $M_2^\prime$ as the test basis and used it to estimate the quantum bit error rate. Where $Q_{1,1;M_1^\prime}=\mu_a \mu_b e^{-\mu_a -\mu_b }Y_{1,1;M_1^\prime}$, and $\frac{1}{2}$ is the basis sifting factor. In the simulation, the gain and error rate are given by:
\begin{center}
$$Y_{1,1;M_1^\prime}=(1-p_d)^2[\frac{\eta_a \eta_b}{2}+(2\eta_a+2\eta_b-3\eta_a\eta_b)p_d+4(1-\eta_a)(1-\eta_b)p_d^2],\eqno{(3)}$$
$$e_{1,1;M_1^\prime}=e_0 Y_{1,1;M_1^\prime}-(e_0-e_d)(1-p_d^2)\frac{\eta_a \eta_b}{2},\eqno{(4)}$$\\
\end{center}
Where $e_0$ is the response error rate caused by the dark count, with $e_0=\frac{1}{2}$, and $e_d$ is the misalignment error rate caused by the phase reference mismatch; $p_d$  is the dark count rate.
\begin{center}
$$Q_{M_1^\prime}=Q_{M_1^\prime}^{(D_0)}+Q_{M_1^\prime}^{(D_1)},\eqno{(5)}$$\\
$$Q_{M_1^\prime}^{(D_0)}=2(1-p_d)^2 e^{-\frac{\mu^\prime}{2}}[1-(1-p_d)e^{-\frac{\eta_a \mu_a}{2}}]\times[1-(1-p_d)e^{-\frac{\eta_b \mu_b}{2}}],\eqno{(6)}$$\\
$$Q_{M_1^\prime}^{(D_1)}=2p_d(1-p_d)^2e^{-\frac{\mu^\prime}{2}}[I_0(2x)-(1-p_d)e^{-\frac{\mu^\prime}{2}}],\eqno{(7)}$$\\
\end{center}
Where $I_0(x)$ is the first type of modified Bessel function. For small values of the variable $x$ a first-order approximation can be used to approximate equation (7).
\begin{center}
$$E_{M_1^\prime} Q_{M_1^\prime}=e_d Q_{M_1^\prime}^{(D_0)}+(1-e_d)Q_{M_1^\prime}^{(D_1)},\eqno{(8)}$$
\end{center}
Here,
\begin{center}
$$\mu^\prime=\eta_a \mu_a+\eta_b \mu_b,\eqno{(9)}$$
$$x=\frac{1}{2}\sqrt{\eta_a \mu_a\eta_b \mu_b},\eqno{(10)}$$
\end{center}
$\eta_a =\eta_b =\frac{\eta}{2}$,$\mu_a=\mu_b=\frac{\mu}{2}$,
Where $\mu^\prime$ represents the average number of photons arriving at the Eve beam splitter. $Q_{M_1^\prime}$ and $E_{M_1^\prime}$ respectively represent the gain of the basis and the error rate of the qubit (ie, $Q_{M_1^\prime}=\sum_{n,m}Q_{n,m;M_1^\prime},E_{M_1^\prime}=\sum_{n,m}Q_{M_1^\prime} e_{n,m;M_1^\prime}/Q_{M_1^\prime}$, $f(E_{M_1^\prime})>1$ is an inefficient function of the error correction process. And $H(x)=-x\log x-(1-x)\log((1-x))$ is the binary Shannon entropy function.
\subsection{Simulation formula of BB84 protocol}
The key rate formula of the decoy BB84 protocol is given in the Ref [7]:

$$R_{BB84}=\frac{1}{2}Q_\mu\{-fH(E_\mu)+q_1[1-H(e_1)]\},\eqno{(11)}$$
Where $\frac{1}{2}$ is the basis sifting factor. In the simulation, the yield and error rate of the $k$-photon component are given by Ref [13]:$$Y_k=1-(1-Y_0)(1-\eta)^k,\eqno{(12)}$$
$$e_k=e_d+\frac{(e_0-e_d)Y_0}{Y_k},\eqno{(13)}$$ The gain and qubit error rate are given by:
$$Q_\mu=\sum_{k=0}^\infty\frac{\mu^k e^{-\mu}}{k!}Y_k=1-(1-Y_0)e^{-\eta\mu},\eqno{(14)}$$
$$E_\mu=\sum_{k=0}^\infty\frac{\mu^k e^{-\mu}}{k!}e_kY_k=e_d+\frac{(e_0-e_d)Y_0}{Q_\mu},\eqno{(15)}$$
Where $Y_0=2p_d$.

\Table{\label{tabl1} Parameters used for simulation
 .{\tt }  {\tt }  }
\br
Parameters&\centre{2}{Values\qquad}\\
\mr

Dark count rate $p_d$&$8\times10^{-8}$\\
Error correction efficiency $f$&1.15\\
Misalignment error $e_d$&$1.5\%$\\
Detector efficiency  $\eta_d$&$14.5\%$\\
\br

\end{tabular}

\end{indented}
\end{table}
\subsection{Comparison of the two protocols}
In this section, we obtain the simulation(Figure 3) of BB84 protocol and our scheme in terms of key rate with formulas in Sections 4.1 and 4.2. It can be seen from the image that the polarization matching scheme based on phase matching exceeds 300 km, and the transmission distance is larger than the transmission distance of the decoy BB84 protocol when the key rate limit is $10^(-15)$. And the polarization scheme has the advantage of not being attacked by the detector side channel. From the perspective of the light source, the polarization scheme uses the weak coherent pulse source, and the BB84 protocol uses a single photon source. From this, it can be seen that our scheme is easy to implement in practice. In the BB84 protocol, the basis used is Z or X basis, and our scheme uses bases that is not unique, can be multiple sets of non-orthogonal bases, and is more general. The selection of the polarization scheme basis is shown in Figure 2. Whenever one value is taken, the polarization scheme has a set of bases.
\begin{figure}[htb!]
 \centering
 \includegraphics[width=13.19cm,height=9.09cm]{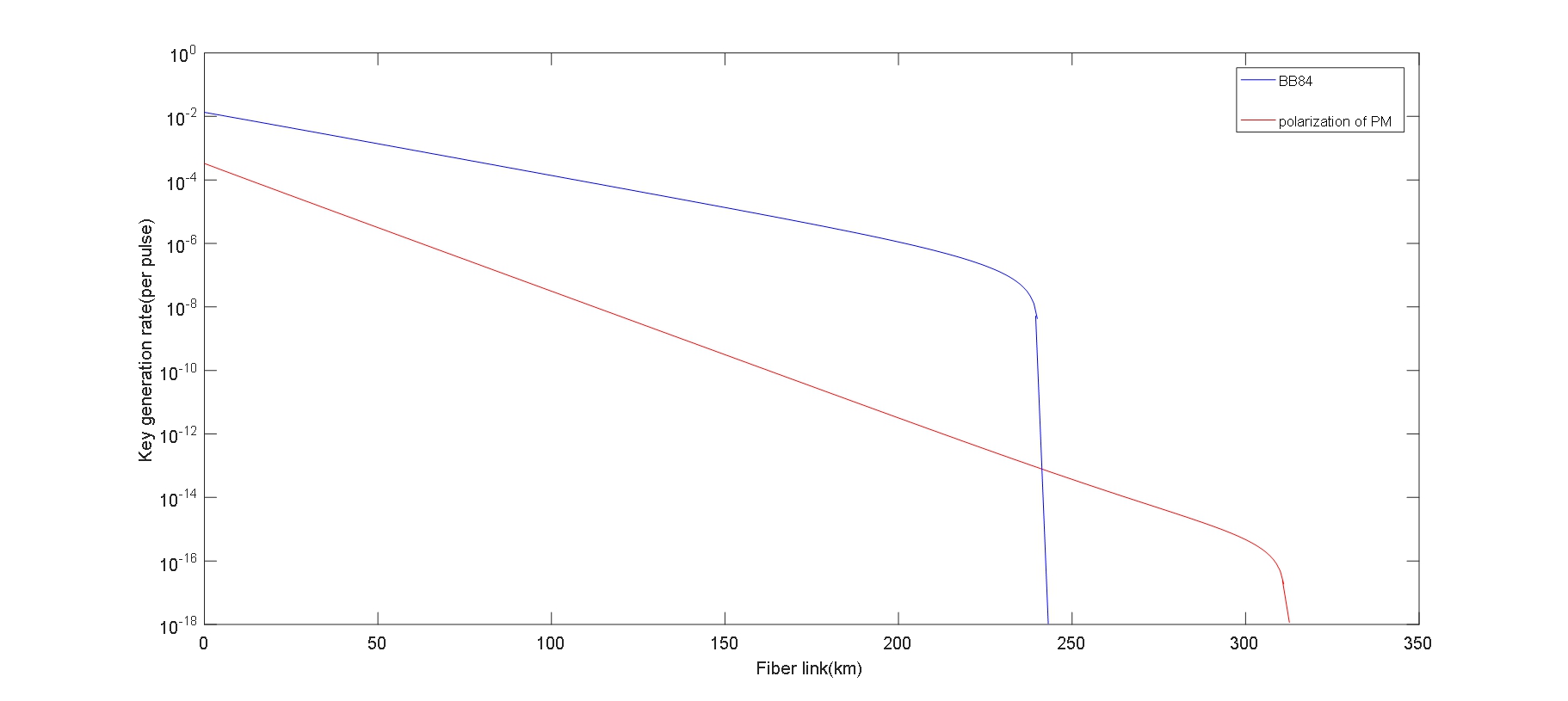}
 \caption{\label{figureone} A simulation of BB84 protocol and our scheme with key rate. }
\end{figure}
\section{Summary}
  Based on the PM-QKD protocol, the polarization scheme of the PM-QKD protocol is given. Also,the key rate formula of its polarization scheme is given. The simulation results show that our scheme is superior to the BB84 protocol in the case of a certain key rate, and it has great advantages in the transmission distance and the selection method of the basis. Our scheme has the advantage of not being attacked by any detectors, and its key rate is also consistent with the MDI-QKD protocol. However, its key rate is not as high as the original protocol. How to increase the key rate is still a problem.


\section*{References}
\begin{thebibliography}{10}
\bibitem{ref1} C. H. Bennett and G. Brassard 1984 Quantum Cryptography:Public Key Distribution and Coin Tossing \textit{IEEE International Conference on Computers}

\bibitem{ref2} C. H. Bennett 1992 Quantum Cryptography Using Any Two Nonorthogonal States  \textit{Phys. Rev. Lett.}$\mathbf{68}$ 3121-24

\bibitem{ref3} Brassard G, Lutkenhaus N, Mor T, et al. 2000 Limitations on practical quantum cryptography  \textit{Phys. Rev. Lett.}$\mathbf{85}$ 1330-33
\bibitem{ref4} Zhao Y, Fung C H F, Qi B, et al. 2008 Quantum hacking: Experimental demonstration of time-shift attack against practical quantum-key-distribution systems  \textit{Phys. Rev. A.}$\mathbf{78}$ 042333
\bibitem{ref5}Makarov V, Skaar J 2007 Faked states attack using detector efficiency mismatch on SARG04, phase-time, DPSK, and Ekert protocols  \textit{Quantum. Inf. Comput.}$\mathbf{8}$ 622
\bibitem{ref6} Makarov V 2012 Controlling passively quenched single photon detectors by bright light  \textit{New J Phys}$\mathbf{11}$ 065003

\bibitem{ref7} X.B.Wang 2005 Beating the photon-number-splitting attack in practical quantum cryptography \textit{Phys. Rev. Lett.}$94$ 230503
\bibitem{ref8} Lopes M, Sarwade N 2018 Optimized decoy state QKD for underwater free space communication  \textit{Int. J. Quantum. Inf.}$\mathbf{16}$ 204-238
\bibitem{ref9}Yu, Z.-W., Zhou, Y.-H. $\&$ Wang, X.-B. 2016 Reexamination of decoy-state quantum key distribution
with biased bases \textit{Phys. Rev. A.}$\mathbf{93}$ 032307
\bibitem{ref10} H.-K. Lo, M. Curty, and B. Qi 2012 Measurement-Device-Independent Quantum Key Distribution  \textit{Phys. Rev. Lett.}$\mathbf{108}$ 130503
\bibitem{ref11}M. Lucamarini, Z. L. Yuan, J. F. Dynes et al. 2018 Overcoming the Rate-Distance Limit of Quantum Key Distribution without Quantum Repeaters  \textit{Nature (London)}$\mathbf{557}$ 400
\bibitem{ref12}Tamaki K , Lo H K , Fung C H F , et al. 2012 Phase encoding schemes for measurement-device-independent quantum key distribution with basis-dependent flaw  \textit{Phys. Rev. A.}$\mathbf{85}$ 042307
\bibitem{ref13} Xiongfeng Ma, Pei Zeng, and Hongyi Zhou 2018 Phase-Matching Quantum Key Distribution  \textit{Phys. Rev. X.}$\mathbf{8}$ 031043
\bibitem{ref14}Namiki R , Hirano T . 2006  Efficient-phase-encoding protocols for continuous-variable quantum key distribution using coherent states and postselection \textit{Phys. Rev. A.}$\mathbf{74}$ 032302
\bibitem{ref15}Pei Zeng, Weijie Wu, and Xiongfeng Ma. 2019 Symmetry-protected privacy: beating the rate-distance linear bound over a noisy channel \textit{arXiv}:1910.05737.
\bibitem{ref16} Choi Y , Kwon O , Woo M , et al. 2016 Plug-and-play measurement-device-independent quantum key distribution \textit{Phys. Rev. A.}$\mathbf{93}$ 032319
\bibitem{ref17}Chun-Hui Zhang, Chun-Mei Zhang, and Qin Wang 2019 Efficient passive measurement-device-independent quantum key distribution  \textit{Phys. Rev. A.}$\mathbf{99}$ 052325
\bibitem{ref18}Ma H X , Huang P , Bai D Y , et al. 2019 Long-distance continuous-variable measurement-device-independent quantum key distribution with discrete modulation  \textit{Phys. Rev. A.}$\mathbf{99}$ 022322
\bibitem{ref19}Wang C , Song X T , Yin Z Q , et al. 2015 Phase-Reference-Free Experiment of Measurement-Device-Independent Quantum Key Distribution  \textit{Phys. Rev. Lett.}$\mathbf{115}$ 160502
\bibitem{ref20}Feihu Xu 2015 Measurement-device-independent quantum communication with an untrusted source \textit{Phys. Rev. A.}$\mathbf{92}$ 012333
\bibitem{ref21}Xiongfeng Ma 2008 Quantum cryptography: from theory to practice  \textit{Toronto:University of Toronto}
\bibitem{ref22}P.W.Shor and J.Preskill 2000 Preskill.Simple Proof of Security of the BB84 Quantum Key Distribution Protocol  \textit{Phys. Rev. Lett.}$\mathbf{85}$
\bibitem{ref23}Ma X , Razavi M 2012 Alternative schemes for measurement-device-independent quantum key distribution  \textit{Phys. Rev. A.}$\mathbf{86}$ 3818-3821
\end{thebibliography}
\end{document}